# Emergence of the Isotropic Kitaev Honeycomb Lattice with Two-dimensional Ising Universality in α-RuCl$_3$

S.-Y. Park[1], S.-H. Do[1,2], K.-Y. Choi[2], D. Jang[3], T.-H. Jang[1,4], J. Schefer[5], C.-M. Wu[6], J. S. Gardner[7], J. M. S. Park[8], J.-H. Park[1,4,9*] and Sungdae Ji[1,4*]

[1]Max Planck POSTECH Center for Complex Phase Materials, Pohang University of Science and Technology, Pohang 37673, Republic of Korea.

[2]Department of Physics, Chung-Ang University, Seoul 06911, Republic of Korea.

[3]Max Planck Institute for Chemical Physics in Solid, 01187 Dresden, Germany.

[4]Department of Physics, Pohang University of Science and Technology, Pohang 37673, Republic of Korea.

[5]Laboratory for Neutron scattering and Imaging (LNS), Paul Scherrer Institut, Villigen PSI CH-5232, Switzerland.

[6]National Synchrotron Radiation Research Center, Hsinchu 30076, Taiwan.

[7]Bragg Institute, Australian Nuclear Science and Technology Organisation, Lucas Heights, New South Wales 2234, Australia.

[8]Neutron Instrumentation Division, Korea Atomic Energy Research Institute, Daejeon 34057, Republic of Korea.

[9]Division of Advanced Materials Science, Pohang University of Science and Technology, Pohang 37673, Republic of Korea.

All the correspondence should be addressed to J.-H.P. (e-mail: jhp@postech.ac.kr) or S.J. (e-mail: sungdae@postech.ac.kr).

**Anderson proposed structural topology in frustrated magnets hosting novel quantum spin liquids (QSLs)[1]. The QSL state is indeed exactly derived by fractionalizing the spin excitation into spinless Majorana fermions in a perfect two dimensional (2D) honeycomb lattice, the so-called Kitaev lattice[2], and its experimental realisation is eagerly being pursued. Here we, for the first time, report the Kitaev lattice stacking with van der Waals (vdW) bonding in a high quality α-RuCl$_3$ crystal using x-ray and neutron diffractions. Even in absence of apparent monoclinic distortion, the system exhibits antiferromagnetic (AFM) ordering below 6.5 K, likely due to minute magnetic interaction from trigonal distortion and/or interlayer coupling additionally to the Kitaev Hamiltonian[3-6]. We also demonstrate 2D Ising-like critical behaviors near the Néel temperature in the order parameter and specific heat, capturing the characteristics of short-range spin-spin correlations underlying the Kitaev model. Our findings hold promise for unveiling enigmatic physics emerging from the Kitaev QSL.**

The seminal work of Anderson triggered a great deal of theoretical and experimental efforts to search for the QSL states in matters, and it has become one of central issues in contemporary condensed matter physics[1]. The QSL state exhibits unconventional features involving quantum fluctuations resulting in absence of magnetic long-range ordering down to zero temperature. One traditional route to realize this novel state has been to explore frustrated magnets such as triangular, kagome, and pyrochlore lattices[7]. On the other hand, Kitaev proposed an exactly solvable model on the ideal 2D honeycomb lattice resulting in topological QSLs and emergent Majorara fermions[2]. The Kitaev model adopts the Ising-like short-range nearest-neighbor (NN) interaction between the ½-spins, which confined in three orthogonal bonds on the honeycomb lattice.

It is known that the 5$d$ transition metal $Ir^{4+}$ ion has a $J_{eff}$ = ½ state due to a strong spin-orbit coupling[8] and its orbital state forms the three orthogonal bonds in the honeycomb lattice to embody the Kitaev model in edge-shared octahedral environments[9]. Until recently, the honeycomb iridates $A_2IrO_3$ ($A$ = Li, Na) have been the focus of intensive research[10-13]. The drawback is monoclinic distortions with anisotropic Ir-Ir bonds, which result in anisotropic magnetic exchange interaction additionally to the Kitaev model Hamiltonian and stabilize a zigzag-type AFM long-range order[3,4].

Quite recently, α-RuCl$_3$ with weak vdW interlayer bonding, which was expected to form a rather ideal 2D honeycomb lattice, has been hailed as a prime candidate material of Kitaev model[14-21]. Indeed, recent Raman and inelastic neutron scattering studies provided signatures of proximate Kitaev QSL behaviors[16,20]. Recent crystallographic studies, however, reported that this layered α-RuCl$_3$ displays a zigzag-type AFM order below $T_N$ ≃ 7 ~ 16 K together with small monoclinic distortion[17,19] in a $C2/m$ space group, controversially to the original report of a trigonal structure with a $P3_112$ space group[22]. The sample dependence of the transition temperature and crystal symmetry casts doubts on whether this ordering and monoclinic distortion are intrinsic to α-RuCl$_3$. Meanwhile, a hysteretic behavior was observable around 150 K in magnetic susceptibility studies[18], suggesting a possibility of a monoclinic to rhombohedral structure transition, as observed in the isostructural CrI$_3$[23], where the honeycomb becomes isotropic with minimal trigonal distortions. Considering that the structural symmetry is vulnerable to stacking faults caused by the weak vdW bonding[19] and the Kitaev interaction is sensitive to the Ru-Ru distance as well as the Ru-Cl-Ru bond angle[4-6], precise structural determination on the high-quality crystal lays a cornerstone for understanding the Kitaev quantum magnetism.

Noticeably, an inelastic x-ray scattering study observed a precursor in the Kitaev model, the bond directional interaction character, in a honeycomb lattice $Na_2IrO_3$ although the AFM order is stabilized in the low temperature preempting the QSL state[24]. In addition, the model requires another key assumption of the Ising-like characters, which can be dictated in the universal asymptotic power-law in the critical exponents[25]. This motivated us to examine thoroughly thermodynamic behaviors near the magnetic phase transition to identify microscopic nature of the spin-spin interaction in α-$RuCl_3$.

Here, we report, for the first time, a clear structural transition of a layered α-$RuCl_3$ using single crystal x-ray and neutron diffractions. The room temperature monoclinic (*C*2/*m*) structure is transformed into the rhombohedral ($R\overline{3}$) structure below 60 K with a true isotropic honeycomb lattice modeled for the Kitaev QSL, nearly free from stacking faults causing the monoclinic distortion. On further cooling, the zigzag-type antiferromagnetic order with **k** = (0 ½ 1) develops below $T_N \simeq 6.5$ K, which is lower than ever reported values ranging from 7 K (single) to 16 K (poly). Remarkably, the critical exponent of the magnetic order parameter was found to be $\beta = 0.11(1)$, close to the theoretical value $\beta = ⅛$ of the 2D Ising model. This critical behavior is further confirmed in the specific heat with the exponent $α = 0$, implying the 2D Ising-like short-range magnetic interaction inherent to the Kitaev model.

Figures 1a and 1b show temperature dependence of the hexagonal lattice constants $a_h$ and $c_h$ estimated by tracking monoclinic $(0\ 6\ 0)_m$ and $(0\ 0\ 4)_m$ x-ray diffraction Bragg peaks on a layered α-$RuCl_3$ single crystal, respectively. The temperature dependence demonstrates a first order structural transition with pronounced thermal hysteresis in a temperature range between $T_{S1} = 166$ K and $T_{S2} = 60$ K. In the in-plane $a_h$, the thermal hysteresis is observable only in a rather narrow temperature range from $T_{S1} = 166$ K to $T^* \simeq 115$ K, while it clearly appears in the $T_{S1}$ to $T_{S2}$ full temperature range in the out-of-plane $c_h$ and a second order type transition behavior becomes apparent at $T^* \simeq 115$ K, indicating existence of three different structural states. Accordingly, we examined the structural symmetry with the reciprocal spacing mapping (RSM) scans on the hexagonal $(h\ k\ 13)_h$ plane at three different temperatures $T$ = 300 K, 80 K, and 4.2 K. For convenient comparisons, we take reciprocal lattice vector transformations of the monoclinic to hexagonal unit cell, given by $\mathbf{a}_m^* = \mathbf{a}_h^* - \mathbf{b}_h^*/2 + \mathbf{c}_h^*$, $2\mathbf{b}_m^* = \mathbf{b}_h^*$ and $\mathbf{c}_m^* = 3\mathbf{c}_h^*$ (see Fig. 1c and Supplementary info 1).

At $T$ = 300 K, the RSM scan shown in Fig. 1d is represented with five dominant reflections of $(1\ 0\ 13)_h$, $(1\ \bar{1}\ 13)_h$, $(1\ 1\ 13)_h$, $(1\ \bar{2}\ 13)_h$ and $(\bar{2}\ 1\ 13)_h$, which are identified to two pairs of monoclinic $(1\ \pm1\ 4)_m$ and $(1\ \pm3\ 4)_m$ reflections, and one $(\bar{2}\ 0\ 5)_m$ reflection, respectively. These reflections satisfy both the reflection condition ($hkl$; $h + k = 2n$) and the $ac$-plane mirror symmetry of the monoclinic space group $C2/m$, as reported previously[17,19]. One can also notice two additional weak reflections at $(\bar{1}\ 0\ 13)_h$ and $(0\ \bar{1}\ 13)_h$, which correspond to 120° twins of the $(1\ \pm1\ 4)_m$ reflections.

The reflection pattern, however, drastically changes below $T_{S2}$. Figure 1e shows the RSM scan at $T$ = 4.2 K. The $(1\ 1\ 13)_h$ and $(1\ \bar{2}\ 13)_h$ reflections corresponding to the $(1\ \pm3\ 4)_m$ ones disappear, and the $(1\ 0\ 13)_h$ 6-fold reflections newly appear, meaning that the low temperature crystal structure is hexagonal, rather than monoclinic. To understand this 6-fold reflection pattern, one can consider two relevant space groups, $P3_112$ and $R\bar{3}$[19,22]. The $P3_112$ space group allows all 6-fold Bragg reflections while only 3-fold ones (yellow circles) are allowed in $R\bar{3}$ but the remaining 3-fold ones is naturally observable due to obverse-reverse twinning in the hexagonal layered crystal[23]. The RSM scan on the warming process (see Fig. 1f) shows only the hexagonal pattern while the scan on the cooling process (see Fig. 2g) does both monoclinic and hexagonal patterns, meaning that structural phase segregation occurs in $T_{S2} < T < T^*$ only upon cooling.

To identify the low temperature crystal structure, we performed single crystal neutron diffraction measurements at $T$ = 5 K, and indexed total 370 peaks in the hexagonal notation. Figure 2a and 2b show the observed structure factor squared $|F_{obs}|^2$ compared with the calculated $|F_{calc}|^2$ for two crystal structure models $R\bar{3}$ and $P3_112$, respectively. The $R\bar{3}$ model, rather than the $P3_112$ one, well explains the diffraction results with very good refinement reliability factors $R_1 = 0.114$ and $wR_2 = 0.250$. In the refinement, the obverse-reverse twin was included and its fraction was estimated to be 0.49(2) implying that the portions of twin and un-twin structures are nearly equal (see details in Supplementary info 2.).

Figure 2c displays the $R\bar{3}$ rhombohedral structure of α-RuCl$_3$ at $T$ = 4 K, determined from the refinement results. The unit cell consists of three honeycomb layers with the interlayer distance 5.64 Å, and each layer stacks with a $(\mathbf{b} - \mathbf{a} + \mathbf{c})/3$ translation. The Ru ions occupy the 6$c$ Wyckoff position $(0, 0, z)$ with a single atomic position variable $z$ in the $R\bar{3}$ space group. The Ru sublattice forms a perfect honeycomb lattice with an equal Ru-Ru bonding length $d_{\text{Ru-Ru}}$ = 3.449 Å at low temperatures ($< T_{\text{S2}}$), differently from the high temperature monoclinic $C2/m$ phase with the anisotropic honeycomb lattice[17,19]. As shown in Fig. 2d, it is noticeable that the Ru-Cl-Ru bond angle in the edge-shared octahedral environment is 94.09°. According to recent theoretical calculations[5,6], the NN Kitaev interaction becomes ferromagnetic (FM) with a near maximum value in a circumstance of a minimal NN Heisenberg one around this angle.

Based on the newly determined low temperature crystal structure, we also examined the magnetic structure in the AFM ground state since the $R\bar{3}$ symmetry only allows zigzag ($\Gamma_1$) and stripe ($\Gamma_2$) magnetic orders, as shown in Fig. 3a and 3b. With the reported magnetic propagation vector $\mathbf{k}$ = (0 ½ 1)$_h$[15,19], the magnetic representation $\Gamma_{\text{mag}}$ is decomposed into these two allowed irreducible representations (IR) $\Gamma_1$ and $\Gamma_2$ for two Ru sites $(0, 0, \pm z)$ with inversion symmetry, and the basis vectors for the magnetic moment in $\Gamma_1$ and $\Gamma_2$ are listed in Table 1. The main difference between these two magnetic structures is the sign of the magnetic exchange coupling between two adjacent interlayer Ru ions, i.e. the AFM (FM) coupling in the $\Gamma_1$ zigzag ($\Gamma_2$ stripe) order. Both magnetic structures are energetically degenerate in the classical Kitaev model of the $R\bar{3}$ symmetry[26], and the ordering type is likely chosen by the additional interlayer coupling.

The magnetic structure at $T$ = 4 K was determined from six magnetic Bragg peaks with the magnetic propagation vector $\mathbf{k}$ = (0 ½ 1)$_h$ obtained from the single crystal neutron diffraction measurements. Figure 3c displays the observed magnetic structure factor squared $|F_{\text{obs}}^{\text{m}}|^2$ compared with the calculated $|F_{\text{calc}}^{\text{m}}|^2$ in the zigzag and stripe magnetic structure models, respectively. The (0 ½ $L$)$_h$ scan, which exhibits four magnetic Bragg peaks, is shown in the inset. $|F_{\text{calc}}^{\text{m}}|^2$ in the zigzag model well reproduces $|F_{\text{obs}}^{\text{m}}|^2$ with much better refinement reliability factors than those in the stripe model (see Supplementary info 3), implying the AFM interlayer coupling in this system. The refined magnetic moment is 0.73(3) $\mu_B$ per Ru and its direction is tilted by 48(3)° from the *ab*-plane, which are significantly larger than the previously reported values of 0.45 $\mu_B$ and 35°[19].

Now, we explore thermodynamic behaviors of the magnetic transition in α-RuCl$_3$. Figure 4a shows the temperature dependence of the magnetic order parameter obtained from integrated intensity of the (0 ½ 1)$_h$ magnetic Bragg peak. By fitting with the power law equation $(1 - T/T_N)^{2\beta}$, the Néel temperature and the critical exponent are determined to be $T_N$ = 6.3(5) K and $\beta$ = 0.11(1), respectively. Remarkably, this $\beta$ value, which is much smaller than that ($\gtrsim$ 0.3) in the 3D Ising or Heisenberg model[25], is close to $\beta$ = ⅛ in the 2D Ising honeycomb model[25]. The 2D Ising criticality is also observable in the magnetic specific heat $C_M$. The inset of Fig. 4b shows the heat capacity $C_p$ and the lattice contribution $C_L$ obtained from an isostructural non-magnetic ScCl$_3$ (see Methods). The $C_M$ is extracted by subtracting $C_L$ from $C_p$, and compared with the theoretical $C_M$ in the 2D Ising honeycomb model with $T_N$ = 6.55 K and a scale factor 0.19 in Fig. 4b. The magnetic entropy release is just 19 % of the expected $R\ln2$ at the transition (see Supplementary info 4), implying presence of an additional magnetic entropy release likely associated with the fractionalization in QSL[27,28]. The small difference in both $T_N$ values is considered due to an experimental error in the neutron measurements. $C_M$ vs $|1 - T/T_N|$ on a logarithmic temperature scale is shown in Fig. 4c. $C_M$ indeed follows the logarithmic diverging behavior, e.g. $\alpha$ = 0, of the 2D Ising honeycomb model by revealing the linear $\log|1 - T/T_N|$ behavior[25] (see Methods). The observed 2D Ising-like critical behaviors suggest that the magnetic transition of α-RuCl$_3$ belongs to the 2D-Ising universality class.

As reported previously[17,19], the crystal structure of α-RuCl$_3$ is monoclinic (*C*2/*m*) in the room temperature. However, we found that the monoclinic structure is transformed into the rhombohedral ($R\bar{3}$) structure. In the previous studies, this structural transition has never been observed and the system remains in the monoclinic structure with anisotropic honeycomb layers in the low temperature, likely due to a considerable amount of stacking faults of the vdW layers in the crystal. In our high quality crystal, the stacking faults become minimized, and the honeycomb lattice becomes isotropic resulting in the rhombohedral structure. Considering the fact that the proximate Kitaev QSL was reported even in the monoclinic phase[16,20], the observation of the rhombohedral structure with the isotropic honeycomb lattice offers a promising playground to study the fractionalized Majorana fermions in the Kitaev QSL physics. Furthermore, the observed pronounced thermal hysteresis suggests an additional temperature scale $T^*$, proposing further investigations in relation with the Majorana fermion excitations[21,27].

Although the zigzag AFM order develops at $T_N \simeq 6.5$ K, which is the lowest temperature ever reported, any symmetry lowering distortion involving magnetoelasticity was not detectable, meaning that the distortion, if present, is even smaller than the experimental resolution (the order of $10^{-3}$ Å). Recent theoretical work in an extended Kitaev-Heisenberg model suggests that the Kitaev interaction becomes either AFM or FM depending on the tilting angle[29]. It is worth to note that the observed 48° tilting angle of the ordered magnetic moment is close to the characteristic angle of 54° for the FM Kitaev interaction in α-RuCl$_3$.

The 2D Ising universality in in α-RuCl$_3$ is in stark contrast to the 3D Ising-type behaviors in Na$_2$IrO$_3$[13], indicating a minimal inter-layer exchange coupling due to the simple vdW bonding involving just the dipolar magnetic interaction. According to theoretical studies on the universality, the XXZ honeycomb compass model has the 3D Ising universality[30], and the 90° compass model does the 2D Ising universality[30]. Meanwhile a theoretical simulation predicted the XY-universality in the Kitaev-Heisenberg model[26]. Given this fact, the 2D Ising universality observed in α-RuCl$_3$ demands future theoretical studies. Nonetheless, the 2D Ising-like behaviors still suggest that the magnetic interaction in α-RuCl$_3$ is close to the NN spin-spin interaction in the Kitaev 2D honeycomb lattice. It is also worth to note that the observed 2D Ising universality in the hexagonal lattice is quite unique. The theoretical 2D Ising model has been realized only in a few square lattice materials (K,Rb)$_2$CoF$_4$[25].

In summary, we presented that the low-temperature crystal structure is identified as the rhombohedral $R\bar{3}$ space group with the perfect honeycomb lattice. We expect that this work stimulates further studies for the effective Hamiltonian involving the Kitaev interaction to describe the ground state as well as the exotic magnetic behaviors in α-RuCl$_3$.

## Methods

### Sample preparation

High quality single crystals of α-RuCl$_3$ were grown by the vacuum sublimation method. A commercial RuCl$_3$ powder (Alfa-Aesar) was thoroughly ground and dehydrated in a vacuum quartz ampule. A sealed ample was placed in a temperature gradient furnace set at 1080 °C. After dwelling for 5 hours, the ample was cooled to 600 °C at the rate of 2 °C per hour. Layered black crystals were obtained at the end of the ample, and the chemical composition was confirmed by using electron dispersive x-ray measurements.

### Single crystal x-ray and neutron diffraction

The single crystal x-ray diffraction was measured by using a Huber four-circle diffractometer equipped with the Rigaku x-ray source (Cu target) in a temperature range 4 K - 300 K. The monochromatic beam at λ = 1.541 Å was produced by a pyrolytic graphite crystal with a resolution of $\Delta d/d \sim 7\times10^{-4}$. The temperature environment was provided by a closed cycle refrigerator with double Be caps. The neutron single crystal crystallographic measurement was carried out at $T$ = 4 K at the TriCS beamline in SINQ, Paul Scherrer Institute. Two incident neutron wavelengths of λ = 1.178 Å and 2.314 Å were utilized to measure the nuclear and magnetic Bragg reflections, respectively. The incident beam resolution is $\Delta d/d \sim 5\times10^{-3}$ for λ = 1.178 Å. The crystal structure was refined with the obverse-reverse twin model[31] by using the SHELX software[32]. The magnetic structure refinement and representation analyses were made by using FullProf Suite[33] and SARA*h*[34], respectively. Temperature dependence of the magnetic peak intensities was obtained in the SIKA in Bragg Institute, Australian Nuclear Science and Technology Organisation by using the cold triple-axis spectrometer. A monochromatic incident neutron beam was set to be at a wavelength of λ = 4.01 Å in a vertical focusing pyrolytic graphite monochromator.

### Heat capacity measurement and 2D Ising model simulation

The heat capacity measurement on a 2.75 mg α-RuCl$_3$ single crystal was carried out by the relaxation method using the Quantum Design Physical Property Measurement System DynaCool. The temperature step was set to be as low as 0.1 K to trace the divergence of heat capacity near $T_N \simeq 6.5$ K. The magnetic heat capacity was estimated by subtracting the lattice contribution, which was obtained from the heat capacity of a 4.2 mg $ScCl_3$ single crystal scaled to the total mass of α-$RuCl_3$.

The exact expression of the magnetic heat capacity in the 2D Ising honeycomb lattice model is calculated from the following equation[35],

$$C_M = k_B K^2 \left\{ -\frac{2}{\sinh^2 2K} + \left(\frac{d\gamma}{dK} - \frac{\gamma}{2\delta}\frac{d\delta}{dK}\right)\mathcal{K}(\delta) + \frac{\gamma}{2\delta(1-\delta)}\frac{d\delta}{dK}\mathcal{E}(\delta) \right\}, \qquad (1)$$

$$\gamma = \begin{cases} \dfrac{(x^4-1)(x^2-4x+1)}{\pi|x^2-1|(x-1)^4} & (T < T_N) \\ \dfrac{(x^4-1)(x^2-4x+1)}{4\pi(x-1)^2\sqrt{x^5-x^4+x^3}} & (T > T_N) \end{cases}, \qquad (2)$$

$$\delta = \begin{cases} \dfrac{16(x^5-x^4+x^3)}{(x^2-1)^2(x-1)^4} & (T < T_N) \\ \dfrac{(x^2-1)^2(x-1)^4}{16(x^5-x^4+x^3)} & (T > T_N) \end{cases}, \qquad (3)$$

where $x = e^{2K}$ with $K = J/k_B T$. $J$ is the exchange interaction between adjacent spins, and $\mathcal{K}$ and $\mathcal{E}$ are elliptic integral of the first and the second kinds. The transition temperature $T_N = 6.55$ K, corresponding to a singular point of $C_M$, is determined in $x = 1/(2-\sqrt{3})$, and $J$ is estimated to be 0.37 meV. Such a small $J$-value is consistent with the 94.09° Ru-Cl-Ru bond angle producing a minimal NN Heisenberg interaction. In comparison with the data, a scale factor, which corresponds to 19% of the expected entropy $R\ln 2$ in the Ising model, was applied to the simulated $C_M$ (see supplementary info 4).

The asymptotic expression of $C_M$ near $\epsilon \equiv 1 - \frac{T}{T_N} = 0$ is given by a form as followings[36],

$$C_M \sim A\frac{|\epsilon|^{-\alpha}-1}{\alpha} + B, \qquad (4)$$

where $A$ and $B$ are constants.

With a mathematical relation, $\lim_{\alpha \to 0} \frac{z^{-\alpha}-1}{\alpha} = \ln z$, the asymptotic expression is reduced to a logarithmic linear function of $\epsilon$ as

$$C_M \sim A\ln|\epsilon| + B. \qquad (5)$$

**References**

1. Anderson, P. W. Resonating valence bonds: A new kind of insulator? *Mater. Res. Bull.* **8,** 153–160 (1973).

2. Kitaev, A. Anyons in an exactly solved model and beyond. *Ann. Phys.* **321,** 2–111 (2006).

3. Rau, J. G., Lee, E. K.-H. & Kee, H.-Y. Generic spin model for the honeycomb iridates beyond the Kitaev limit. *Phys. Rev. Lett.* **112,** 077204 (2014).

4. Winter, S. M., Li, Y., Jeschke, H. O. & Valenti, R. Challenges in design of Kitaev materials: Magnetic interactions from competing energy scales. *Phys. Rev. B* **93,** 214431 (2016).

5. Kim, H.-S. & Kee, H.-Y. Crystal structure and magnetism in α−$RuCl_3$: An ab initio study. *Phys. Rev. B* **93,** 155143 (2016).

6. Yadav, R. *et al.* Spin-orbit excitation energies, anisotropic exchange, and magnetic phases of honeycomb $RuCl_3$. Preprint at http://arxiv.org/abs/1604.04755v1 (2015).

7. Lacroix, C., Mendels, P. & Mila, F. *Introduction to Frustrated Magnetism*. **164,** (Springer, Berlin, 2011).

8. Kim, B. J. *et al.* Novel $J_{eff}$=1/2 Mott state induced by relativistic spin-orbit coupling in $Sr_2IrO_4$. *Phys. Rev. Lett.* **101,** 076402 (2008).

9. Jackeli, G. & Khaliullin, G. Mott insulators in the strong spin-orbit coupling limit: From Heisenberg to a quantum compass and Kitaev models. *Phys. Rev. Lett.* **102,** 017205 (2009).

10. Chaloupka, J., Jackeli, G. & Khaliullin, G. Kitaev-Heisenberg model on a honeycomb lattice: possible exotic phases in iridium oxides $A_2IrO_3$. *Phys. Rev. Lett.* **105,** 027204 (2010).

11. Singh, Y. & Gegenwart, P. Antiferromagnetic Mott insulating state in single crystals of the honeycomb lattice material $Na_2IrO_3$. *Phys. Rev. B* **82,** 064412 (2010).

12. Choi, S. K. *et al.* Spin waves and revised crystal structure of honeycomb iridate $Na_2IrO_3$. *Phys. Rev. Lett.* **108,** 127204 (2012).

13. Ye, F., Chi, S., Cao, H. & Chakoumakos, B. C. Direct evidence of a zigzag spin-chain structure in the honeycomb lattice: A neutron and x-ray diffraction investigation of single-crystal $Na_2IrO_3$. *Phys. Rev. B* **85,** 180403 (2012).

14. Plumb, K. W., Clancy, J. P., Sandilands, L. J. & Shankar, V. V. α−RuCl$_3$: A spin-orbit assisted Mott insulator on a honeycomb lattice. *Phys. Rev. B* **90,** 041112(R) (2014).

15. Sears, J. A. *et al.* Magnetic order in α−RuCl$_3$: A honeycomb-lattice quantum magnet with strong spin-orbit coupling. *Phys. Rev. B* **91,** 144420 (2015).

16. Sandilands, L. J., Tian, Y., Plumb, K. W., Kim, Y.-J. & Burch, K. S. Scattering continuum and possible fractionalized excitations in α-RuCl$_3$. *Phys. Rev. Lett.* **114,** 147201 (2015).

17. Johnson, R. D. *et al.* Monoclinic crystal structure of α-RuCl$_3$ and the zigzag antiferromagnetic ground state. *Phys. Rev. B* **92,** 235119 (2015).

18. Kubota, Y., Tanaka, H., Ono, T., Narumi, Y. & Kindo, K. Successive magnetic phase transitions in α−RuCl$_3$: *XY*-like frustrated magnet on the honeycomb lattice. *Phys. Rev. B* **91,** 094422 (2015).

19. Cao, H. B. *et al.* Low-temperature crystal and magnetic structure of α−RuCl$_3$. *Phys. Rev. B* **93,** 134423 (2016).

20. Banerjee, A. *et al.* Proximate Kitaev quantum spin liquid behaviour in a honeycomb magnet. *Nat. Mater.* **15,** 733–740 (2016).

21. Nasu, J., Knolle, J., Kovrizhin, D. L., Motome, Y. & Moessner, R. Fermionic response from fractionalization in an insulating two-dimensional magnet. *Nat. Phys.* advance online publication, 4 July 2016 (DOI 10.1038/nphys3809).

22. Stroganov, E. V. & Ovchinnikov, K. V. Crystal structure of ruthenium trichloride. *Vestnik. Leningrad. Univ. Ser. Fiz. i Khim* **12,** 22–152 (1957).

23. McGuire, M. A., Dixit, H., Cooper, V. R. & Sales, B. C. Coupling of crystal structure and magnetism in the layered, ferromagnetic insulator CrI$_3$. *Chem. Mater.* **27,** 612–620 (2015).

24. Chun, S. H. *et al.* Direct evidence for dominant bond-directional interactions in a honeycomb lattice iridate Na$_2$IrO$_3$. *Nat. Phys.* **11,** 462–466 (2015).

25. De Jongh, L. J. & Miedema, A. R. Experiments on simple magnetic model systems. *Adv. Phys.* **23,** 1–260 (1974).

26. Price, C. C. & Perkins, N. B. Critical properties of the Kitaev-Heisenberg model. *Phys. Rev. Lett.* **109,** 187201 (2012).

27. Nasu, J., Udagawa, M. & Motome, Y. Thermal fractionalization of quantum spins in a Kitaev model: Temperature-linear specific heat and coherent transport of Majorana fermions. *Phys. Rev. B* **92,** 115122 (2015).

28. Yamaji, Y. *et al.* Clues and criteria for designing a Kitaev spin liquid revealed by thermal and spin excitations of the honeycomb iridate $Na_2IrO_3$. *Phys. Rev. B* **93,** 174425 (2016).

29. Chaloupka, J. & Khaliullin, G. Magnetic anisotropy in the Kitaev model systems $Na_2IrO_3$ and $RuCl_3$. *Phys. Rev. B* **94,** 064435 (2016).

30. Nussinov, Z. & van den Brink, J. Compass models: Theory and physical motivations. *Rev. Mod. Phys.* **87,** 1–59 (2015).

31. Herbst-Irmer, R. & Sheldrick, G. M. Refinement of obverse/reverse twins. *Acta Crystallogr. B* **58,** 477–481 (2002).

32. Sheldrick, G. M. A short history of SHELX. *Acta Crystallogr., A, Found. Crystallogr.* **64,** 112–122 (2008).

33. Rodríguez-Carvajal, J. Recent advances in magnetic structure determination by neutron powder diffraction. *Physica B Condens. Matter.* **192,** 55–69 (1993).

34. Wills, A. S. A new protocol for the determination of magnetic structures using simulated annealing and representational analysis (SARAh). *Physica B Condens. Matter.* **276-278,** 680–681 (2000).

35. Houtappel, R. M. F. Order-disorder in hexagonal lattices. *Physica* **16,** 425–455 (1950).

36. Ikeda, H., Hatta, I. & Tanaka, M. Critical heat capacities of two-dimensional Ising-like antiferromagnets $K_2CoF_4$ and $Rb_2CoF_4$. *J. Phys. Soc. Jpn.* **40,** 334–339 (1976).

**Acknowledgements**

This work is supported by the National Research Foundation (NRF) through the Ministry of Science, ICP & Future Planning (MSIP) (No. 2016K1A4A4A01922028). S.-H.D. and K.-Y.C. are supported by the Korea Research Foundation (KRF) grant (No. 2009-0076079) funded by the Korea government (MEST). J.M.S.P. is supported by the NRF under the contract NRF-2012M2A2A6002461. This work

**Author contributions**

S.J. and K.-Y.C. conceived the work. S.-H.D. and K.-Y.C. synthesized samples. S.-Y.P. and S.J. performed the x-ray diffraction experiments. S.-Y.P., J.M.S.P., and S.J. carried out the neutron diffraction experiments with the support from J.S., C.-M.W. and J.S.G. S.-Y.P. and S.J analyzed the x-ray and neutron data. S.-H.D., S.-Y.P. and T.-H.J. performed the heat capacity measurements. S.-H.D., S.-Y.P. and D.J. analyzed the heat capacity data. S.-Y.P., K.-Y.C. J.-H.P. and S.J. wrote the manuscript with contributions from all coauthors. S.J. and J.-H.P. led the project.

**Competing financial interests**

The authors declare no competing financial interests.

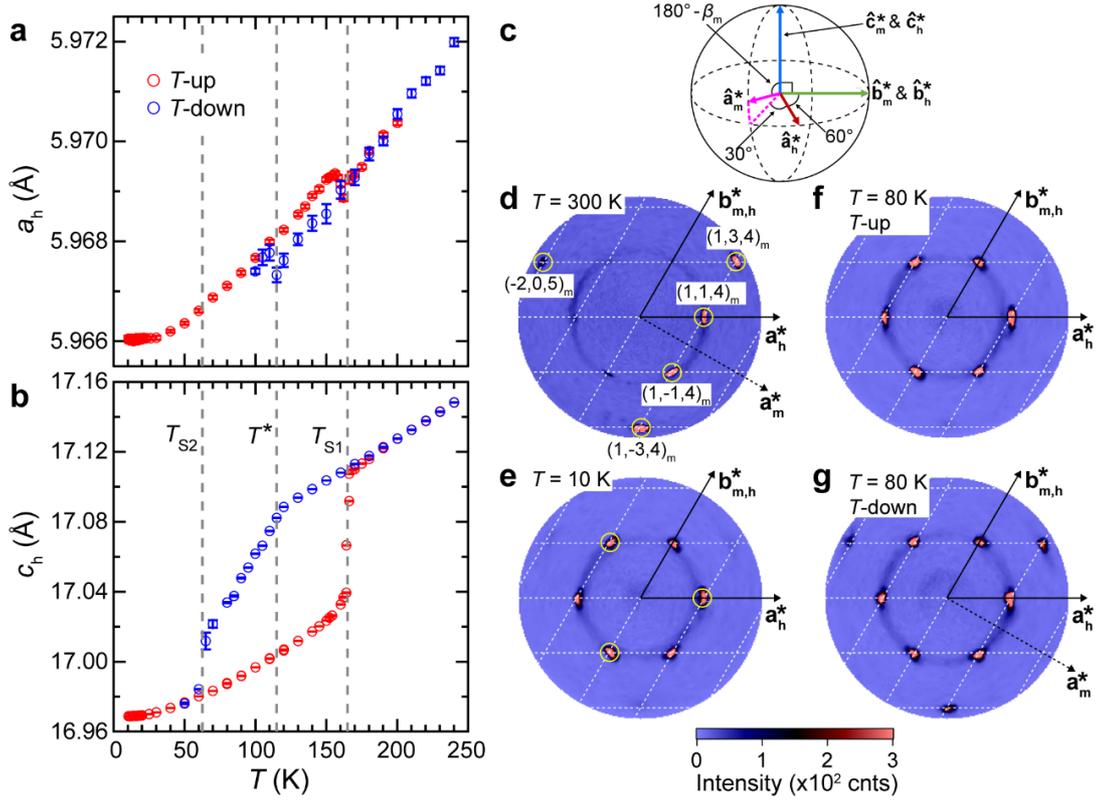

**Figure 1 | Temperature dependent lattice parameters and reciprocal space mapping scans**. **a-b**, Temperature dependent hexagonal lattice parameters (**a**) $a_h$ and (**b**) $c_h$ obtained from single crystal x-ray diffraction measurements on respective monoclinic $(0\ 6\ 0)_m$ and $(0\ 0\ 4)_m$ Bragg peaks. A cooling (warming) process is presented by blue (red) circles. Vertical dashed lines guide two structural transition temperatures, $T_{S1}$ = 166 K and $T_{S2}$ = 62 K, and one inflection temperature $T^* \simeq$ 115 K. **c**, Reciprocal unit vectors in the monoclinic and hexagonal phases. **d**, Reciprocal space mapping (RSM) scans on a hexagonal $(h\ k\ 13)_h$ plane at $T$ = 300 K. The color scale bar represents the diffracted x-ray intensity. White dashed lines guide for the hexagonal reciprocal space in the $(h\ k\ 13)_h$ plane. The RSM scan is dominated by five Bragg peaks (yellow circles) at both temperatures, which satisfy the monoclinic $C2/m$ space group reflection conditions, $hkl$; $h + k = 2n$ and the $ac$-plane mirror symmetry. **e**, RSM scan at $T$ = 10 K. Six-folded Bragg reflection peaks, which are allowed in the $P3_112$ space group, also represents the $R\bar{3}$ space group ($-h + k + l = 3n$, yellow circles) with obverse-reverse twinning ($h - k + l = 3n$). **f-g,** The RSM scans at $T$ = 80 K upon (**f**) warming and (**g**) cooling. In warming, the RSM exhibits the hexagonal pattern while it does both monoclinic and hexagonal patterns in cooling.

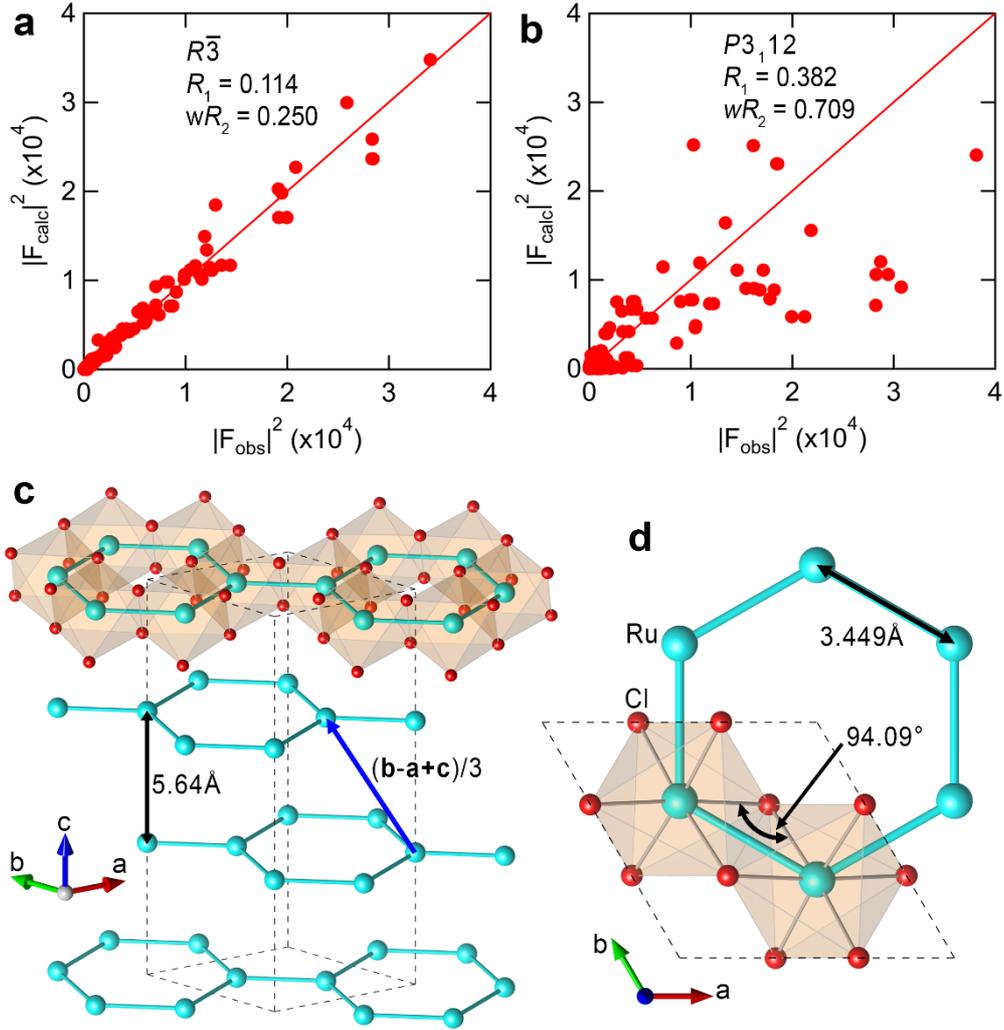

**Figure 2 | Crystal structure and refinements**. **a-b**, Observed structure factor square $|F_{obs}|^2$ versus calculated $|F_{calc}|^2$ for (**a**) $R\bar{3}$ and (**b**) $P3_112$ models. $|F_{obs}|^2$ was obtained from the single crystal neutron diffraction measurements at $T = 4$ K. Red lines present the guide line $|F_{obs}|^2 = |F_{calc}|^2$. **c**, Three dimensional stacking of Ru (cyan spheres) honeycomb layers in the refined rhombohedral $R\bar{3}$ structure. The unit cell (dashed lines) consists of three Ru honeycomb layers with the interlayer distance 5.64 Å, and each Ru layer is sandwiched by two Cl layers (red spheres) and stacks with a $(\mathbf{b} - \mathbf{a} + \mathbf{c})/3$ translation (blue arrow). Edge-shared RuCl$_6$ octahedra form the Ru honeycomb layer. **d**, A Ru$_6$ hexagon with two selected adjacent edge-shared RuCl$_6$ octahedra. In the $R\bar{3}$ structure, Ru ions form a perfect honeycomb lattice with an equal Ru-Ru bond length $d_{\text{Ru-Ru}} = 3.449$ Å. All the Ru-Cl-Ru bond angles are 94.09° in the edge-shared RuCl$_6$ octahedra.

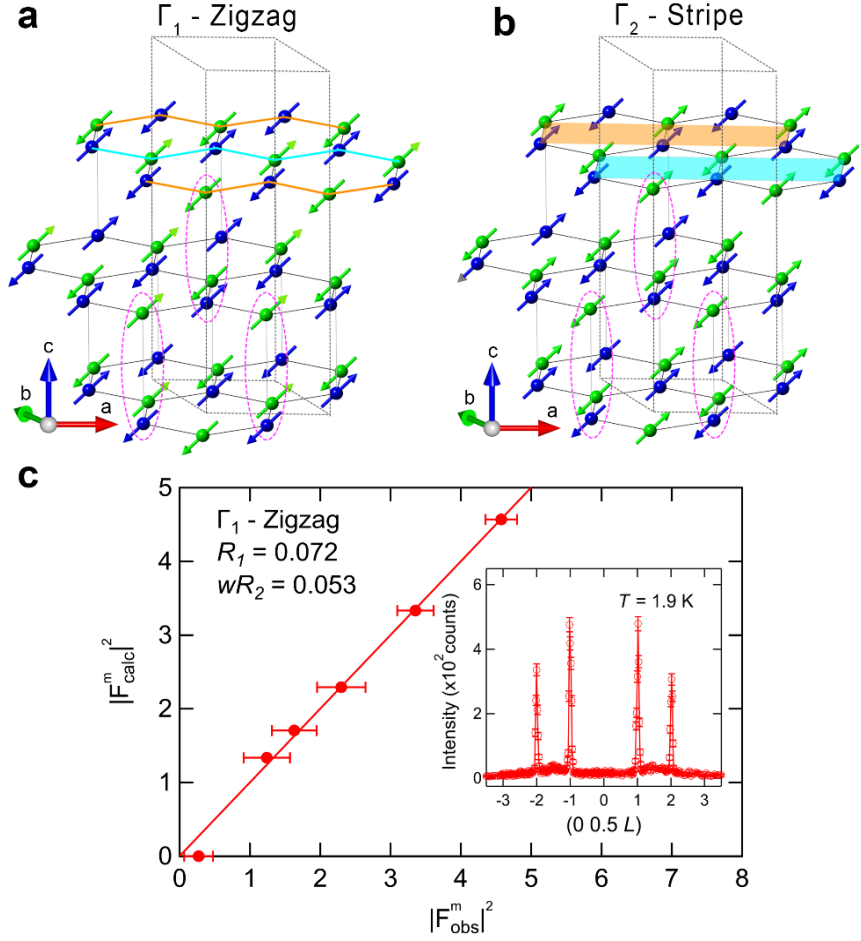

**Figure 3 | Magnetic structure and refinement**. **a-b**, Schematic diagrams of (**a**) zigzag and (**b**) stripe antiferromagnetic (AFM) structures obtained from the representational analysis of the $R\bar{3}$ space group with a magnetic propagation vector **k** = (0 ½ 1). The magnetic representation $\Gamma_{mag}$ is decomposed into two allowed irreducible representations $\Gamma_1$ (zigzag) and $\Gamma_2$ (stripe) for two Ru$_1$ (0, 0, $z$) and Ru$_2$ (0, 0, $\bar{z}$) sublattices with an inversion symmetry. Blue (green) spheres and arrows denote the Ru$_1$ (Ru$_2$) sites and their magnetic moment directions, respectively. The AFM and FM couplings between two adjacent inter-layer Ru$_1$ and Ru$_2$ ions are recognized from the magnetic orders in elliptical red dotted lines for the $\Gamma_1$ (**a**) and $\Gamma_2$ (**b**) representations, respectively. Grey dotted lines represent the unit cell of the $R\bar{3}$ structure. **c**, Observed structure factor square $|F_{obs}^m|^2$ versus calculated $|F_{calc}^m|^2$ in the zigzag model. $|F_{obs}^m|^2$ was obtained from the single crystal neutron diffractions at $T$ = 4 K. A red line presents $|F_{obs}^m|^2$ =$|F_{calc}^m|^2$. The inset shows a line scan along (0 ½ $L$), which is dominated by two pairs of magnetic Bragg peaks at $L$ = ±1 and ±2, confirming the magnetic propagation vector. A red line is a guide for eyes.

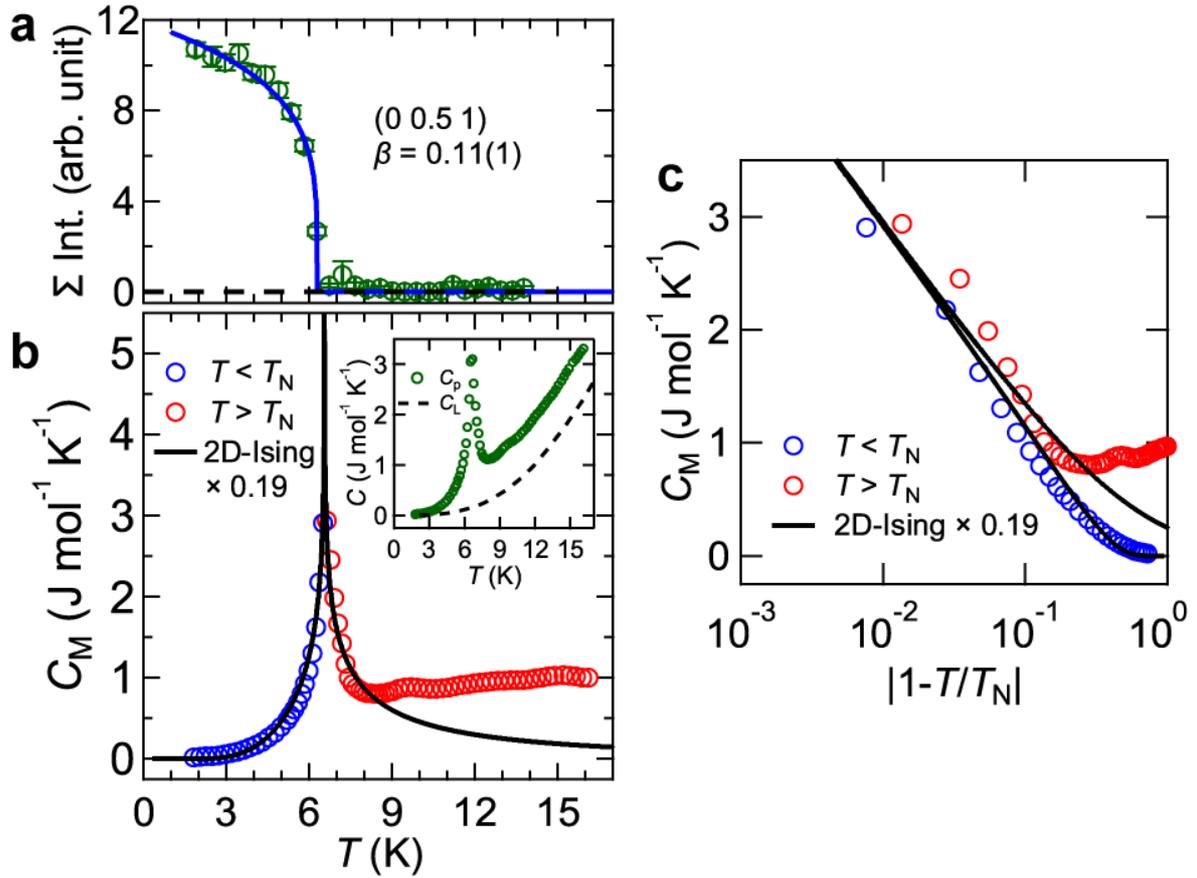

**Figure 4 | Magnetic order parameter and specific heat**. **a**, Temperature dependence of the integrated intensity of a magnetic Bragg peak (0 ½ 1) obtained from the neutron diffraction measurements on α-RuCl$_3$ single crystals. The blue solid line represents the critical exponent fit with $(1 - T/T_N)^{2\beta}$, where $T_N$ = 6.3(5) K and $\beta$ = 0.11(1). **b**, Magnetic heat capacity $C_M$ of α-RuCl$_3$ (red and blue circles) and scaled 2D Ising honeycomb model fit (solid line). The inset shows the heat capacity $C_p$ (green circles) of α-RuCl$_3$ and the lattice contribution $C_L$ (dashed line). $C_p$ also exhibits the magnetic transition at $T_N$ = 6.55 K. The $C_M$ is extracted from subtracting $C_L$ from $C_p$ (see methods). The red and blue circles represent $C_M$ above and below $T_N$, respectively. **c**, $C_M$ versus $|1 - T/T_N|$ in a logarithmic temperature scale. $C_M$ follows the logarithmic divergence behavior, e.g. $\alpha$ = 0, in the 2D Ising honeycomb model by revealing the linear $\log|1 - T/T_N|$ behavior. The small difference in $T_N$ is attributed to a temperature probing error in the neutron measurements.

**Table 1** | Basis vectors $\psi_i$ of irreducible representations $\Gamma_1$ and $\Gamma_2$ obtained from a representational analysis for the space group $R\bar{3}$ with a magnetic propagation vector **k** = (0 ½ 1) of the Ru$_1$ (0 0 $z$) and Ru$_2$ (0 0 $\bar{z}$) sublattices ($z$ = 0.3333) in the three hexagonal crystallographic bases **a$_h$**, **b$_h$** and **c$_h$**.

|  | $\Gamma_1$ (zigzag) | | | $\Gamma_2$ (stripe) | | |
|---|---|---|---|---|---|---|
| site | $\psi_1$ | $\psi_2$ | $\psi_3$ | $\psi_4$ | $\psi_5$ | $\psi_6$ |
| Ru$_1$ (0 0 $z$) | (1 0 0) | (0 1 0) | (0 0 1) | (1 0 0) | (0 1 0) | (0 0 1) |
| Ru$_2$ (0 0 $\bar{z}$) | ($\bar{1}$ 0 0) | (0 $\bar{1}$ 0) | (0 0 $\bar{1}$) | (1 0 0) | (0 1 0) | (0 0 1) |

# Emergence of the Isotropic Kitaev Honeycomb Lattice with Two-dimensional Ising Universality in α-RuCl$_3$

## Supplementary Info 1. Reciprocal lattice vector transformation

In order to obtain the reciprocal lattice vector transformation between a monoclinic and a hexagonal structure, we make the corresponding unit cell vector transformation between two structures,

$$\begin{cases} \mathbf{a}_m = \mathbf{a}_h \\ \mathbf{b}_m = \mathbf{a}_h + 2\mathbf{b}_h \\ \mathbf{c}_m = (-\mathbf{a}_h + \mathbf{c}_h)/3 \end{cases}, \quad (1)$$

as shown in Supplementary Fig. 1. The monoclinic unit cell volume $V_m$ is related to the hexagonal one $V_h$ with $3V_m = 2V_h$.

The reciprocal lattice vectors of monoclinic and hexagonal unit cells can be written as

$$\begin{cases} \mathbf{a}_m^* = \dfrac{\mathbf{b}_m \times \mathbf{c}_m}{V_m} \\ \mathbf{b}_m^* = \dfrac{\mathbf{c}_m \times \mathbf{a}_m}{V_m} \\ \mathbf{c}_m^* = \dfrac{\mathbf{a}_m \times \mathbf{b}_m}{V_m} \end{cases} \text{ and } \begin{cases} \mathbf{a}_h^* = \dfrac{\mathbf{b}_h \times \mathbf{c}_h}{V_h} \\ \mathbf{b}_h^* = \dfrac{\mathbf{c}_h \times \mathbf{a}_h}{V_h} \\ \mathbf{c}_h^* = \dfrac{\mathbf{a}_h \times \mathbf{b}_h}{V_h} \end{cases}, \quad (2)$$

and $\mathbf{a}_m^*$'s are expressed with $\mathbf{a}_h^*$ using Supplementary Eqn. (1) as follows,

$$\mathbf{a}_m^* = \frac{\mathbf{b}_m \times \mathbf{c}_m}{V_m} = \frac{(\mathbf{a}_h + 2\mathbf{b}_h) \times (-\mathbf{a}_h + \mathbf{c}_h)}{3V_m} = \frac{1}{2V_h}[2(\mathbf{b}_h \times \mathbf{c}_h) - (\mathbf{c}_h \times \mathbf{a}_h) + 2(\mathbf{a}_h \times \mathbf{b}_h)]$$

$$= \mathbf{a}_h^* - \frac{1}{2}\mathbf{b}_h^* + \mathbf{c}_h^*,$$

$$\mathbf{b}_m^* = \frac{\mathbf{c}_m \times \mathbf{a}_m}{V_m} = \frac{(-\mathbf{a}_h + 2\mathbf{b}_h) \times \mathbf{a}_h}{3V_m} = \frac{\mathbf{c}_h \times \mathbf{a}_h}{2V_h} = \frac{1}{2}\mathbf{b}_h^*,$$

$$\mathbf{c}_m^* = \frac{\mathbf{a}_m \times \mathbf{b}_m}{V_m} = \frac{\mathbf{a}_h \times (\mathbf{a}_h + 2\mathbf{b}_h)}{V_m} = 3\frac{\mathbf{c}_h \times \mathbf{a}_h}{V_h} = 3\mathbf{c}_h^*. \quad (3)$$

The monoclinic reflection indices $(h\ k\ l)_m$ are transformed to the hexagonal $(h\ k\ l)_h$ by

$$h_m \mathbf{a}_m^* + k_m \mathbf{b}_m^* + l_m \mathbf{c}_m^* = h_m(\mathbf{a}_h^* - \mathbf{b}_h^*/2 + \mathbf{c}_h^*) + k_m(\mathbf{b}_h^*/2) + l_m(3\mathbf{c}_h^*)$$

$$= h_m \mathbf{a}_h^* + (-h_m/2 + k_m/2)\mathbf{b}_h^* + 3l_m \mathbf{c}_h^* = h_h \mathbf{a}_h^* + k_h \mathbf{b}_h^* + l_h \mathbf{c}_h^*, \quad (4)$$

and expressed as a matrix form,

$$\begin{pmatrix} h_m \\ k_m \\ l_m \end{pmatrix} = \begin{pmatrix} 1 & 0 & 0 \\ 1 & 2 & 0 \\ -1/3 & 0 & 1/3 \end{pmatrix} \begin{pmatrix} h_h \\ k_h \\ l_h \end{pmatrix}. \quad (5)$$

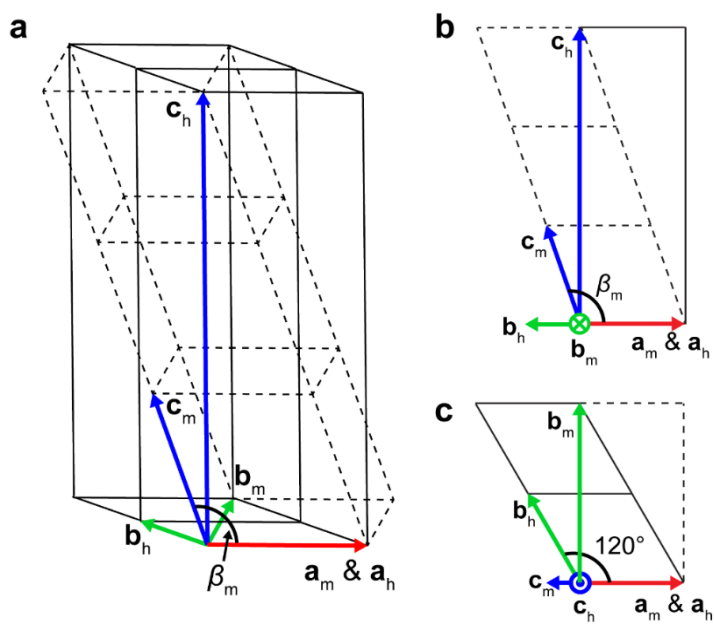

Supplementary Figure 1. **a**, schematic diagrams of hexagonal unit cell vectors $a_h$'s and monoclinic ones $a_m$'s. The hexagonal and monoclinic unit cells are represented by solid and dashed lines, respectively. **b-c**, The projection of both unit cells to (**b**) *ac*- and (**c**) *ab*-plane.

## Supplementary Info 2. Crystal structure refinement and parameters

The rhombohedral $R\bar{3}$ structure was determined from refinements of total 370 Bragg peaks collected by the single crystal neutron diffraction at $T = 4$ K. In the refinements, we include its reverse-obverse twins which commonly occur in the rhombohedral space group. In the hexagonal setting for the $R\bar{3}$ space group, the reflection conditions for the obverse and reverse domains are represented as $-h + k + l = 3n$ and $h - k + l = 3n$, respectively. All Bragg peaks well satisfied the obverse/reverse reflection conditions. We provide refined crystallographic parameters of the rhombohedral $R\bar{3}$ structure as listed in the Supplementary Table 1. Refined lattice parameters are $a = b = 5.973(1)$ Å and $c = 16.93(6)$ Å in the hexagonal setting and the reverse to obverse twin ratio is 0.48(2) to 0.52(2).

Supplementary Table 1. Structural parameters refined by a space group $R\bar{3}$ with a reverse-observe twin model for data obtained from single crystal neutron diffraction measurements at $T = 5$ K. Refined hexagonal lattice parameters are $a = b = 5.973(1)$ Å and $c = 16.93(6)$ Å. Reliability factors are $R_1 = 0.114$ and $wR_2 = 0.250$ with a reverse-observe twin ratio of 0.48(2) : 0.52 (2).

|  | Ru | Cl |
|---|---|---|
| Site | 6*c* | 18*f* |
| *x* | 0 | 0.31786(45) |
| *y* | 0 | 0.33413(42) |
| *z* | 0.33330(28) | 0.41181(12) |
| $U_{eq}$ | 0.0164(11) | 0.0163(7) |

# Supplementary Info 3. Comparison of magnetic structure refinements for the zigzag and the stripe model.

The magnetic representation analysis for the determined $R\bar{3}$ space group allows two zigzag ($\Gamma_1$) and stripe ($\Gamma_2$) magnetic structures with the propagation vector **k** = (0 1/2 1) as listed in the Table 1 of the main text. In this supplementary, we compare refinement results for both magnetic structures. Supplementary Figure 3 shows squares of the experimental magnetic structure factors versus the theoretical values for the zigzag and stripe spin orderings. As a result, the experimental values well agree with the theoretical ones for the zigzag structure with much better refinement reliability factors of $R_1 = 0.072$ and $wR_2 = 0.053$ in comparison with $R_1 = 0.136$ and $wR_2 = 0.179$ for the stripe structure.

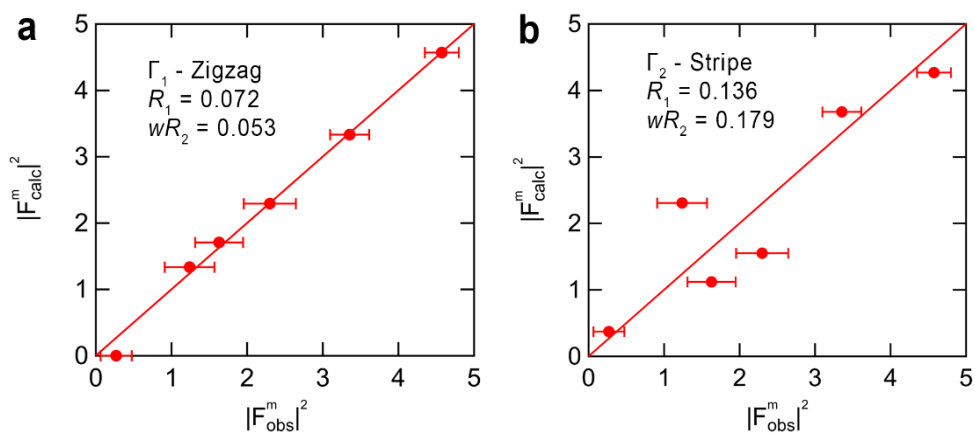

Supplementary Figure 2. Squares of the experimental magnetic structure factors versus the theoretical values for the (a) zigzag ($\Gamma_1$) and (b) stripe ($\Gamma_2$) magnetic structure models.

# Supplementary Info 4. Partial magnetic entropy release at $T_N$

Supplementary Figure 4a shows the magnetic specific heat $C_M$ of α-RuCl$_3$ in comparison with the theoretical $C_M$ for the 2D Ising honeycomb model with $T_N$ = 6.55 K multiplied by a scaling factor 0.19. The theoretical $C_M$ well agrees with the observed one below $T$ < 8 K. The observed $C_M$, however, exhibits a broad tail above $T$ > 8 K, implying presence of certain surviving magnetic correlations in the long-range Ising-like spin order. Moreover, the released entropy $S_M$, which is obtained by integrating the $C_M$ versus $T$, is estimated to be only 19% of the value $R$ln2 at expected for ordering s = ½ moments at the transition, and it keep increasing above $T$ >$T_N$, as shown in Supplementary Fig. 4b. It is noticed that such results are indeed expected in the thermal fractionalization of a spin-1/2 state predicted in the Kitaev QSL [1,2], although the further experimental confirmations for the fractionalization are necessary.

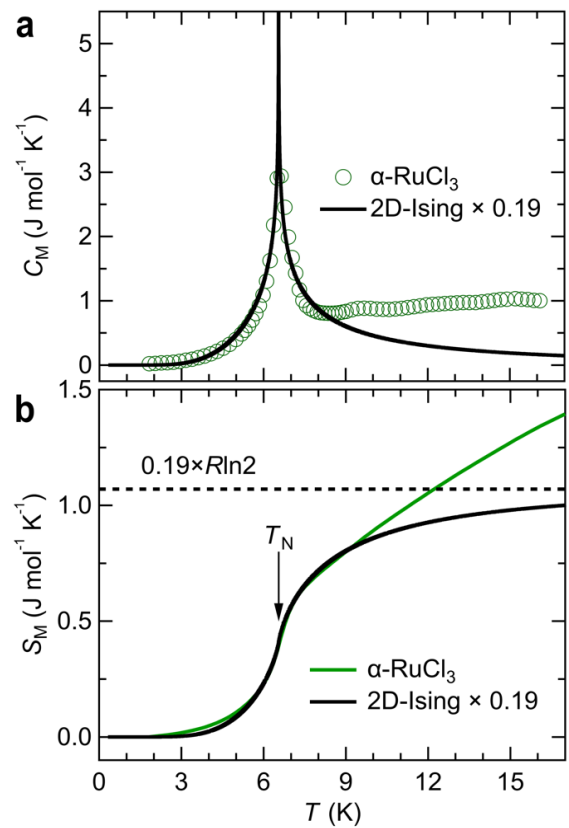

Supplementary Figure 3. **a**, Magnetic heat capacity $C_M$ (green circles) of α-RuCl$_3$ compared with the theoretical $C_M$ (black solid line) in a 2D Ising honeycomb model with $T_N$ = 6.55 K multiplied by a scale factor 0.19. **b**, Magnetic entropy $S_M$ obtained from integration of $C_M$ over $T$. The green and black solid line represents $S_M$'s of α-RuCl$_3$ and the model, respectively.

**Supplementary References**

[1] Nasu, J., Udagawa, M. & Motome, Y. Thermal fractionalization of quantum spins in a Kitaev model: Temperature-linear specific heat and coherent transport of Majorana fermions. *Phys. Rev. B* **92,** 115122 (2015)

[2] Yamaji, Y. *et al*. Clues and criteria for designing a Kitaev spin liquid revealed by thermal and spin excitations of the honeycomb iridate Na$_2$IrO$_3$, *Phys. Rev. B* **93**, 174424 (2016).